# *In vivo* whole-cell recording from morphologically identified mouse superior colliculus neurons

**Robin Broersen[1,3,4*] and Greg J. Stuart[1,2]**
[1]Eccles Institute of Neuroscience, John Curtin School of Medical Research, Australian National University, Canberra, ACT, Australia
[2]Current affiliation: Department of Physiology, Monash University, Melbourne, VIC, Australia
[3]Current affiliation: Department of Neuroscience, Erasmus MC, 3000 CA, Rotterdam, The Netherlands
[4]Lead contact
[*]Correspondence: r.broersen@erasmusmc.nl

## Summary

*In vivo* whole-cell recording when combined with morphological characterization after biocytin labeling is a powerful technique to study subthreshold synaptic processing in cell-type-identified neuronal populations. Here, we provide a step-by-step procedure for performing whole-cell recordings in the superior colliculus of urethane-anesthetized mice, a major visual processing region in the rodent brain. Two types of visual stimulation methods are described. While we focus on superior colliculus neurons, this protocol is applicable to other brain areas.

## Highlights

- Whole-cell recordings from neurons in the visual layers of the superior colliculus in urethane-anesthetized mice.
- Full-field or complex visual stimulus methods enable studying visual processing at the synaptic and spike level.
- Recorded neurons are labeled for cell-type identification and morphology.
- Combined physiological and morphological for structure-function encoding in cell-type identified neurons.

## Before you begin

### Preparing anesthesia mixture, recording and immunostaining solutions

**Timing: 4 h**

1. Prepare urethane-chlorprothixene (CPX) mixture for anesthesia (see materials and equipment).
    a. A mixture of urethane (dose: 1 g/kg) and CPX (dose: 5 mg/kg) is administered by intraperitoneal injection. Aliquots can be stored at -20°C for several months. To make a 50 mL stock concentration:
        i. Fill a 50 mL falcon tube with 25 mL sterile injectable saline solution.
        ii. Add 5 g urethane and 0.05 g chlorprothixene hydrochloride to the solution.
        iii. Fill up the tube to 50 mL with saline solution and vortex the solution until the chemicals are dissolved.

iv. Aliquot the solution in 1 mL volumes and store aliquoted vials in the -20°C freezer.

**CRITICAL: Urethane and CPX are toxic. Work in a laboratory safety cabinet and wear appropriate personal protective equipment (PPE).**

2. Prepare artificial cerebrospinal fluid (aCSF) (see materials and equipment).
   a. aCSF is used during recordings to keep the brain surface moist. We use a standard recipe for aCSF (Haider et al., 2016), but various aCSF recipes can be used. New aCSF needs to be prepared every two weeks if stored at 4°C. To make 500 mL of aCSF:
      i. Fill a 500 mL flask with 300 mL Millipore water and add a magnetic stirrer bar.
      ii. Add 3.945 g NaCl, 0.201 g KCl, 0.596 g HEPES, 0.102 g $MgCl_2.6H_2O$, and 0.1 g $CaCl_2$.
      iii. Stir until dissolved.
      iv. Measure and set the pH to 7.3 using 1M KOH.
      v. Top up to 500 mL with Millipore water.

3. Prepare intracellular solution (ICS) for the recording electrode (see materials and equipment).
   a. This potassium gluconate-based ICS can be prepared, aliquoted and stored at -20°C for several months. This is a standard recipe in our laboratory and has been used in published work (Longordo et al., 2013). Work on ice to preserve the quality of the chemicals. To make 50 mL of ICS:
      i. Fill a 50 mL Falcon tube with 25 mL Millipore water
      ii. Add 0.037 g KCl, 1.522 g K-Gluconate, 0.119 g HEPES, 0.101 g Mg-ATP, 0.008 g $Na_2$-GTP and 0.191 g $Na_2$-phosphocreatine.
      iii. Add Millipore water until the volume is 40 mL. Stir until the salts are fully dissolved.
      iv. Measure the pH and set to 7.25-7.35.
      v. Measure the osmolarity and add Millipore water until the volume is 50 mL with an osmolarity of 290-300 mOsm.
      vi. Aliquot 0.75 mL into 1.5 mL Eppendorf tubes and store at -20°C.

4. Prepare phosphate-buffered saline (PBS) for immunostaining (see materials and equipment).
   a. 1X PBS is used during immunostaining and for preparing paraformaldehyde fixative (see below). A 10X stock solution can be prepared, stored at 4°C for several weeks and diluted as required. To make 1 L of 10X PBS:
      i. Fill a 1 L flask with 800 mL Millipore water and add a magnetic stirrer bar.
      ii. Add 80 g NaCl, 2 g KCl, 13.3 g $Na_2HPO_4.2H_2O$ and 2.4 g $KH_2PO_4$
      iii. Stir until chemicals are dissolved.
      iv. Measure and set pH to 7.4 with HCl. Fill up to 1 L with Millipore water. Store at 4°C.
      v. For 1X PBS dilute 100 mL of 10X PBS in 900 mL of Millipore water and store at 4°C.

5. Prepare 4% paraformaldehyde (PFA) solution for transcardial perfusion (see materials and equipment).

a. PFA solution should ideally be made fresh every week. To make 1 L of 4% PFA solution:
              i. Fil a glass bottle with 800 mL of 1X PBS warmed to ~60 °C and add a magnetic stirrer bar. PBS can be heated up in a microwave, but avoid boiling. Place on a heated stir plate in a laboratory safety cabinet.
             ii. Add 40 g of PFA powder.
            iii. Add a few drops of 1 M NaOH.
            iv. Stir until PFA is completely dissolved.
             v. With paper pH indicators measure and set pH to 7.4 with HCl. Fill up to 1 L with 1X PBS.
            vi. Let the solution cool down sufficiently and store at 4°C.

**CRITICAL: PFA is toxic. Work in a laboratory safety cabinet and wear appropriate PPE.**

   6. Prepare soap mixture for shaving (see materials and equipment).
        a. We use a mixture of regular hand soap and Millipore water to shave the hairs on top of the head, because in our experience this gives a better shaving result than using a hair trimmer and the loose hairs can be more easily removed. A tube with soapy water can be stored at 4°C for two weeks.
              i. Fill a 50 mL falcon tube with 45 mL Millipore water.
             ii. Add ~1 mL hand soap and mix well. Shake carefully to prevent formation of foam as much as possible. Store at 4°C.

## Preparation before the experiment

**Timing: 30 min**

1. Prepare patch electrodes (1-1.5 μm tip diameter, 4-7 MΩ resistance) from filamented borosilicate glass pipettes using a flaming/brown micropipette puller.
    a. We use a Sutter Micropipette Puller (Model P-97) with a 4-stage pull protocol to create glass electrodes for *in vivo* whole-cell recordings.
          i. Position glass pipetted (1.5 mm outer diameter, 0.86 mm inner diameter) in the puller and start the pull program.
         ii. Inspect one of the two pipettes after each pull under the microscope (400x magnification) to detect any abnormalities. Immediately discard any patch electrodes that have an unusual tip diameter, show signs of damage or are contaminated with debris.

**Optional:** Tip resistance estimate. To estimate the tip resistance connect the patch electrode to a piece of plastic tubing attached to a 10 mL syringe with the plunger positioned at the 10 mL mark. Submerge the tip of the electrode in a 80% ethanol alcohol solution and press the plunger to build up pressure inside the electrode. Air bubbles should start appearing when the plunger reaches between the 7 mL (earlier indicates too low tip resistance) and 5.5 mL mark (later indicates too high tip resistance). This is a quick way to test the resistance of the glass electrode.

**Note**: Refer to the Sutter Instrument Company 'Pipette Cookbook', available on their [website](), for instructions on how to optimize the pulling program. We recommend making new glass electrodes every day before the recording session and not to reuse pipettes from earlier days. Store pipettes in a covered container.

**Note**: We recommend making at least 8 patch electrodes to ensure smooth continuation of the experiment once the brain is accessible.

2. Prepare the solutions for the experiment.
    a. Prepare a syringe with ICS with biocytin for back-filling whole-cell glass electrodes.
        i. Thaw a tube containing 0.75 mL ICS (if frozen).
        ii. Add 0.5% (w/v) biocytin to the ICS (3.75 mg per 0.75 mL ICS). Vortex the solution until all biocytin has been dissolved.
        iii. Fill a 1 mL syringe with the tube contents. Position a 0.22 µm syringe filter on the end of the syringe and attach a microloader tip. Keep on ice until needed next to the patching rig.
    b. Prepare a syringe with urethane-CPX mixture for general anesthesia.
        i. Thaw a tube containing 1 mL urethane-CPX mixture (if frozen).
        ii. Fill a 1 mL syringe with the tube contents and attach a 25G injection needle.
        iii. Leave in the laboratory safety cabinet until needed.
    c. Prepare a syringe with bupivacaine for local anesthesia.
        i. Fill a 1 mL syringe with 0.2 mL bupivacaine hydrochloride solution (5 mg/ml).
        ii. Attach a 30G injection needle.
    d. Prepare a syringe with aCSF.
        i. Fill a 10 mL syringe with aCSF and attach a 18G injection needle.
    e. Prepare a syringe with diluted pentobarbitone sodium (Lethabarb) for euthanasia.
        i. Dilute stock solution of pentobarbitone (325 mg/mL) to 60 mg/mL. For 10 mL total volume add 1.85 mL pentobarbitone with 8.15 mL sterile saline solution.
        ii. Fill a 1 mL syringe with 0.1 mL diluted pentobarbitone solution.
        iii. Attach a 25G injection needle.

**CRITICAL: The urethane-CPX mixture is toxic. Work in a laboratory safety cabinet and wear appropriate personal protective equipment (PPE). Avoid contact with the skin.**

**Note**: If the microloader tip does not fit on the syrine filter a shortened 20 µL pipette tip can be used as a connection between the filter and the tip.

3. Prepare the surgery table/area. Sterilize the surgery area with ethanol. It is important to minimize the distance between the surgery area and the experimental rig to reduce cooling of the mouse during transport. It is recommended that all surgical procedures are performed on a feedback-controlled heating pad (37°C).
    a. Cover the heating pad with a tissue to collect any fluids generated during surgery. Secure the covered heating pad to the table using tape. Place the following items next to the surgery area:
        i. A small weighing boat filled with soapy water.
        ii. UV light, optibond primer and light-curing dental composite.
        iii. Several fluid-absorbing swaps.
        iv. 10 mL syringe filled with aCSF.
        v. A Pasteur pipette with 0.2 mL silicon oil.
    b. Sterilize the surgery tools. The following tools are needed for the surgery:
        i. Scalpel holder with blade (curved, no. 22), dissecting scissors, surgical curette and forceps (dumont #3).

  ii. Use 80% ethanol alcohol solution and a tissue to clean and sterilize surgical tools. Place on a tissue next to the surgery area.

# Key resources table

| REAGENT or RESOURCE | SOURCE | IDENTIFIER |
|---|---|---|
| Antibodies | | |
| Vectastain Elite ABC-HRP Kit | Vector Laboratories | Cat# PK-6100 |
| Chemicals, Peptides, and Recombinant Proteins | | |
| Urethane | Sigma-Aldrich | Cat# U2500; CAS:51-79-6 |
| Chlorprothixene hydrochloride | Sigma-Aldrich | Cat# C1671; CAS: 6469-93-8 |
| Sterile injectable saline | Alpha Medical Solutions | Cat# BBR3521340-P |
| Paraformaldehyde | Sigma-Aldrich | Cat# 158127; CAS: 30525-89-4 |
| Biocytin | Sigma-Aldrich | Cat# B4261; CAS: 576-19-2 |
| Hydrogen peroxide (30%) | VWR | Cat# BDH7690-1; CAS: 7722-84-1 |
| Triton X-100 | Sigma-Aldrich | Cat# X100; CAS: 9036-19-5 |
| Bovine serum albumin (BSA) | VWR | Cat# 421501J; CAS: 9048-46-8 |
| **Experimental Models: Organisms/Strains** | | |
| Mouse: C57BL/6J (male/female, 4-6 weeks old) | The Jackson Laboratory | Strain# 00064; RRID: IMSR_JAX:000664 |
| **Software and Algorithms** | | |
| AxoGraph | AxoGraph | Version: AxoGraph 1.7.6 |
| MATLAB | MathWorks | Version: R2012b |
| Fiji | ImageJ2 | Release 2.5.0 |
| Illustrator | Adobe | Version: 25.1 |
| Other | | |
| Borosilicate glass (filamented, fire polished) | Sutter Instruments | Cat# BF150-86-7.5HP |
| MILLEX GV Syringe Filter (0.22 μm) | Sigma-Aldrich | Cat# SLGV013SL |
| Microloader Tips, Eppendorf | VWR | Cat# 89009-310 |
| BD Microlance 18G needle | VWR | Cat# 613-3833 |
| BD Microlance 25G needle | VWR | Cat# 613-0902 |
| BD Microlance 30G needle | VWR | Cat# 613-3942 |
| OptiBond Universal - Bottle Kit | Kerr Dental | Cat# 36517 |
| Kulzer Venus Flow composite | DC Dental | Cat# 504-66014561 |
| Absorption Triangles - Unmounted | Fine Science Tools | Cat# 18105-03 |
| Scalpel blades No. 22 | Fisher Scientific | Cat# 22-079-697 |
| Dissecting scissors - L 4 1/2 in., Sharp, straight | Sigma-Aldrich | Cat# Z265977 |

| Surgical micro curette | Fine Science Tools | Cat# 10082-15 |
| Dumont #3 Forceps | Fine Science Tools | Cat# 11231-30 |
| Dumont #5/45 Forceps | Fine Science Tools | Cat# 11251-35 |
| Dumont #7 Forceps - Curved | Fine Science Tools | Cat# 11274-20 |
| Graefe forceps - Curved 1.2 x 0.9 mm | Fine Science Tools | Cat# 11652-10 |
| VectaMount Permanent Mounting Medium | Vector Laboratories | Cat# H-5000-60 |
| Tissue-Tek Cryomold | Sakura | Cat# 4566 |
| Tissue-Tek OCT compound | Sakura | Cat# 4583 |

# Materials and equipment

### Urethane-chlorprothixene (CPX) mixture

| Reagent | Final concentration (mg/mL) | Amount |
| --- | --- | --- |
| Urethane | 20 | 1 g |
| Chlorprothixene hydrochloride | 1 | 0.05 g |
| Sterile saline solution | n/a | Add to 50 mL |

Aliquots can be stored at -20°C for several months.

### Artificial cerebrospinal fluid (aCSF) (Haider et al., 2016)

| Reagent | Final concentration (mM) | Amount |
| --- | --- | --- |
| NaCl | 135 | 3.945 g |
| KCl | 5.4 | 0.201 g |
| HEPES | 5 | 0.596 g |
| $MgCl_2 \cdot 6H_2O$ | 1 | 0.102 g |
| $CaCl_2$ | 1.8 | 0.1 g |
| Millipore water | n/a | Add to 500 mL |

Can be stored in sealed container at 4°C for two weeks.

### Intracellular solution (ICS)

| Reagent | Final concentration (mM) | Amount |
| --- | --- | --- |
| KCl | 10 | 0.037 g |
| K-gluconate | 130 | 1.522 g |
| HEPES | 10 | 0.119 g |
| Mg-ATP | 4 | 0.101 g |
| $Na_2$-GTP | 0.3 | 0.008 g |
| $Na_2$-phosphocreatine | 15 | 0.191 g |
| Millipore water | n/a | Add to 50 mL |

Aliquots can be stored at -20°C for several months.

### 10X phosphate-buffered saline (PBS)

| Reagent | Final concentration (mM) | Amount |
| --- | --- | --- |
| NaCl | 1370 | 80 g |
| KCl | 27 | 2 g |
| $Na_2HPO_4 \cdot 2H_2O$ | 100 | 14.4 g |

| KH$_2$PO$_4$ | 18 | 2.4 g |
|---|---|---|
| Millipore water | n/a | Add to 1 L |

Can be stored in sealed container at 4°C for several weeks.

**1 M NaOH solution**

| Reagent | Final concentration (M) | Amount |
|---|---|---|
| NaOH | 1 | 4 g |
| Millipore water | n/a | Add to 100 mL |

Can be stored at -20°C for several months.

**1 M HCl solution**

| Reagent | Final concentration (M) | Amount |
|---|---|---|
| HCl (37% solution) | 1 | 8.2 |
| Millipore water | n/a | Add to 100 mL |

Can be stored at -20°C for several months.

**4% paraformaldehyde (PFA) solution**

| Reagent | Final concentration | Amount |
|---|---|---|
| Paraformaldehyde | 4% | 40 g |
| 1 M NaOH | n/a | Few drops |
| 1 M HCl | n/a | Few drops |
| 1X PBS | n/a | Add to 1 L |

Can be stored in sealed container at 4°C for one week.

**Soap mixture for shaving**

| Reagent | Final concentration | Amount |
|---|---|---|
| Standard hand soap | n/a | 1 mL |
| Millipore water | n/a | Add to 50 mL |

Can be stored at -4°C for one week.

**30% sucrose solution**

| Reagent | Final concentration | Amount |
|---|---|---|
| Sucrose | 30% | 30 g |
| 1X PBS | n/a | Add to 100 mL |

Can be stored at -4°C for two weeks.

**Pentobarbitone sodium solution**

| Reagent | Final concentration (mg/mL) | Amount |
|---|---|---|
| Pentobarbitone sodium (325 mg/mL) | 60 | 1.85 mL |
| Sterile saline solution | n/a | Add to 10 mL |

Can be stored at -4°C for two weeks

**1% H$_2$O$_2$ solution**

| Reagent | Final concentration | Amount |
|---|---|---|
| H$_2$O$_2$ (30% solution) | 1% | 166.7 μL |
| 1X PBS | n/a | Add to 5 mL |

Mix well and use immediately.

**BSA and triton X-100 solution**

| Reagent | Final concentration | Amount |
|---|---|---|
| BSA | 1% | 50 mg |
| Triton X-100 | 0.3% | 15 µL |
| 1X PBS | n/a | Add to 5 mL |

Mix well and use immediately.

**ABC solution (Vectastain Elite ABC kit)**

| Reagent | Final concentration | Amount |
|---|---|---|
| BSA | 1% | 50 mg |
| Triton X-100 | 0.3% | 15 µL |
| 1X PBS | n/a | Add to 5 mL |
| Solution A | n/a | 1 drop |
| Solution B | n/a | 1 drop |

Mix well and use immediately.

# Step-by-step method details

## Preparation of the animal

**Timing: 45 min**

This part involves preparing the animal for *in vivo* whole-cell recordings. The mouse is anesthetized, a head bar is secured on the skull and a craniotomy is then performed.

**Note**: All experimental procedures are carried out in accordance with the Australian code for the care and use of animals for scientific purposes (8th ed., 2013, NHMRC). All experiments have been approved by the Animal Experimentation Ethics Committee of the Australian National University (ANU), Canberra, Australia. The exact protocol used should be first approved by the animal ethics committee of your research institution.

1. Anesthetize the animal. **[Troubleshooting 1](#)**
    a. Weigh the animal and determine the urethane-CPX mixture injection volume:
        i. Injection volume = weight of mouse in grams x 0.01 mL of urethane-CPX mixture. For a 20 g mouse, 0.2 mL urethane-CPX mixture should be injected intraperitoneally (IP). Final dosage: 1 g/kg urethane and 5 mg/kg CPX.
    b. Determine the bupivacaine injection volume.
        i. Injection volume = weight of mouse in grams x 1.4 µL bupivacaine hydrochloride (5 mg/mL). For a 20 g mouse, 0.028 mL should be injected subcutaneously. Final dosage: 6-8 mg/kg
    c. Place the mouse cage on a separate, constant voltage heating platform.
    d. Prefill a gas anesthesia induction box with isoflurane (4-5% in 0.3 L/min $O^2$).
    e. Place a C57Bl/6J mouse in the induction box and monitor its breathing.
    f. Remove the mouse from the induction box once breathing has slowed down. Assess the depth of anesthesia using the paw withdrawal reflex.
        i. Place the mouse on a tissue on the surgery table.
        ii. Carefully pinch one of the back paws with forceps (Dumont #3).

iii. If the mouse is not sufficiently anesthetized, it will show a withdrawal reflex. In that case, place the mouse back in the induction box.
  g. If no withdrawal reflex is detected, inject the mouse IP with the predetermined volume of urethane-CPX mixture for long-term anesthesia.
    i. Place the mouse back in its (pre-heated) cage and wait for the anesthesia to take effect.
  h. After 15 minutes, assess the animal's depth of anesthesia using the paw withdrawal reflex (step 1f).
    i. Administer additional urethane-CPX mixture if necessary (max 10% of initial dose).

**Note**: Reassess the paw withdrawal reflex every 5 minutes during the preparation of the animal to make sure that sufficiently deep anesthesia is maintained. Administer additional urethane-CPX mixture if necessary (max 10% of initial dose).

2. Attach the head bar and prepare the skull for craniotomy. **[Troubleshooting 2](#)**
   a. Place the anesthetized mouse on the heating pad, with the head facing the researcher (Figure 1A). Level the head by putting an object under the snout of the mouse (e.g. an absorbent triangle), so that Bregma and Lambda are approximately at equal height.
     i. Cover the temperature sensor with vaseline and place the probe in the rectum. Secure the sensor to the table using tape. Make sure that the temperature of the mouse is kept at 37°C.
   b. Apply a drop of silicon oil on both eyes to prevent the eyes drying out.
   c. Trim the whiskers on both sides.
   d. Remove the hair on top of the head using the scalpel dipped in soapy water. Keep the scalpel and hairs wet, as this makes the shaving easier.
     i. Softly press on the skin to aid the shaving. Work from the neck forward to the snout, hold the scalpel in a 30° angle to the head surface and make small shaving movements. Remove the loose hairs with a tissue.

**Alternative:** An electric hair trimmer or hair removal cream can be used.

**Note**: Take care not to damage the eyes, since they are critical to the experiment. The short time between induction of anesthesia and covering the eye with silicon oil would normally not lead to drying out of the eyes. If signs of drying of the eyes emerge (e.g. white cornea), immediately apply silicon oil on the eyes.

   e. Dry the shaved area with an absorbent triangle.
   f. Inject bupivacaine subcutaneously under the skin on top of the head for local anesthesia. Wait >5 minutes to allow the anesthesia to take effect.
   g. Remove the skin covering the top of the skull (Figure 1B).
     i. Using forceps (#3) and dissection scissors, cut from the front to back of the head starting from the skin between the eyes to the back of the skull at the level of the ears. The skin covering the Lambda skull landmark should be removed.
     ii. Remove the remaining tissue from the skull bone using a surgical curette.
     iii. Clean the bone with absorbent triangles and a drop of aCSF if necessary. Wait until any bleeding from the bone has stopped and the bone is fully dry before proceeding to the next step.

h. Apply a layer of Optibond primer on the skull bone, covering the skull between the eyes up until the Bregma landmark (Figure 1C).
      i. Let the primer etch the bone for 10 seconds.
      ii. Apply air to dry and harden using UV light for 20 seconds.
      iii. Repeat this step one more time.
   i. Secure the head bar to the skull.
      i. Apply a layer of UV-curing dental flow composite on top of the primer (Figure 1D).
      ii. Place the head bar on top of the dental composite and harden using UV light for 20 seconds.

**Note:** Make sure to attach the head bar straight. Also make sure that the composite or head bar does not occlude the area where the craniotomy is made. The head should be level and the head bar should be parallel to the eyes (Figure 1E).

      iii. Apply dental composite around and on top of the head bar to increase contact with the skull. Harden using UV light for 20 seconds.
   j. Place the mouse in the electrophysiology rig.
      i. Remove the rectal probe
      ii. Position the mouse on a heating pad in the recording rig and secure the head bar. Insert the new rectal probe and turn on the heating pad control unit. Maintain the animal temperature at 37°C.

**Pause point**: After placing the mouse in the electrophysiology rig and making sure that the body temperature of the mouse is stable, one can take a short break. Regularly check the depth of anesthesia (every 15 min) and administer additional urethane-CPX mixture if necessary (max 10% of initial dose).

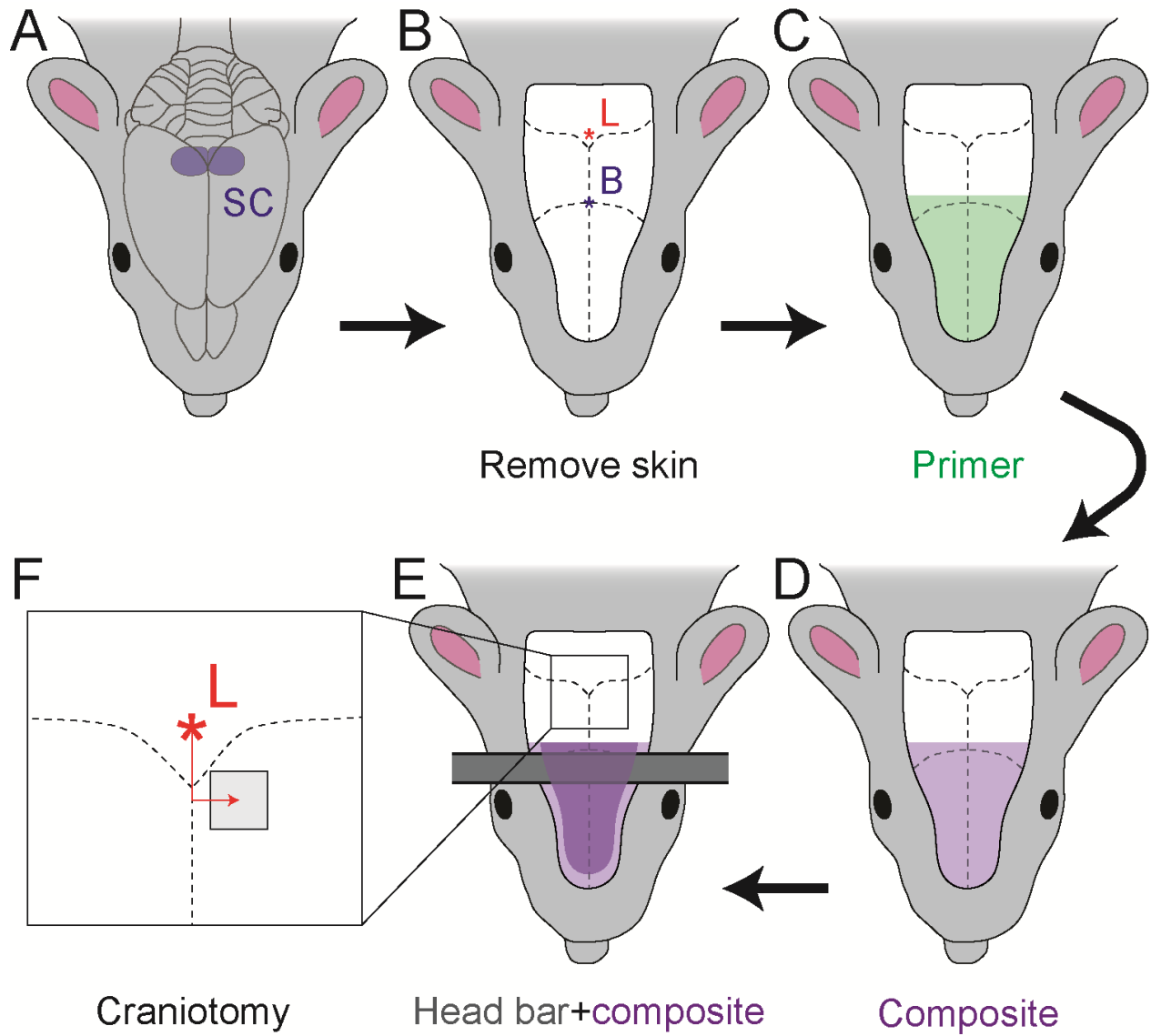

**Figure 1.** Head bar placement and craniotomy. Schematic representation of steps taken for head bar placement and the craniotomy surgery. Anesthetized mice are shaved and skin is removed, exposing the skull where Bregma (B) and Lambda (L) landmarks can be identified. Two layers of primer are applied on the dry skull and hardened with UV light. A layer of UV-curing composite is used to attach the head bar to the skull, which is secured with a second layer of composite. The position of the craniotomy is determined based on coordinates relative to Lambda (L).

## Performing the craniotomy surgery

**Timing: 20 min**

At this point a craniotomy is made so that the SC can be accessed with recording electrodes. The dura mater covering the brain surface will also be removed.

3. Determine the coordinates of the center of the craniotomy and indicate the location on the skull.
    a. Attach a glass electrode to the pipette holder, which is positioned perpendicular to the horizontal plane.
        i. Lower the tip of the electrode to right above the Lambda landmark and set the micromanipulator coordinates to '0' (see red asterisk in Figure 1F).
        ii. Using the appropriate stereotaxic coordinates position the tip of the electrode above the SC (see below).
        iii. Make a scratch in the skull at the position of the electrode tip using a fine needle to mark the site of the craniotomy.

**Note:** A glass electrodes from a previous recording session can be used to determine the site of the craniotomy to preserve newly made electrodes.

**Note:** The SC can be targeted between -1.1 mm and +0.5 mm in the anteroposterior axis relative to Lambda. If using Bregma as a landmark, the SC can be targeted between -3.1 mm and -4.7 mm in the anteroposterior axis relative to Bregma. These distances assume an adult mouse with a typical Bregma-Lambda distance of 4.21 mm. In the mediolateral axis, the SC can be targeted between 0 and 1.75 mm relative to the midline, depending on the anteroposterior position. For recordings of binocular visual neurons, targeting the anteromedial part of SC is recommended (0.8 mm to 1.0 mm anterior to Lambda, 0.1 mm to 0.8 mm lateral to midline).

4. Make the craniotomy and remove the dura.
    a. Use a dental drill with a small bit size (<0.5 mm Ø) to create a 1 x 1 mm craniotomy. We perform a square craniotomy, since it is easier to drill four straight tracks than to drill in a circular motion. It is also easier to monitor the thickness of the skull of a straight drill track. Together this improves the quality of the craniotomy.
        i. Make four superficial drill tracks around the determined coordinates.
        ii. Frequently apply aCSF to prevent the heat produced by the drill to reach the brain tissue. Dry using absorbent triangles before drilling again.

**CRITICAL: Make sure to apply very light pressure on the drill and to not touch the underlying brain tissue with the drill head. A craniotomy of optimal quality without damage or bleeding in the underlying brain tissue is crucial for the success of the experiment. Putting in enough time and effort to create an optimal craniotomy will pay off in the form of good quality recordings.**

       iii. Stop the drilling process once the skull under the four drill tracks is equally thin. Ideally the skull in the drill track should start to show small cracks. Softly touching the piece of skull covering the craniotomy should easily cause the piece to move.
       iv. With forceps (#5/45 bended tip) carefully grab one corner of the piece of skull to be removed and pull it off slowly.

**Note:** Occasionally the piece of skull and dura mater do not easily separate. Do not pull off the skull piece if this happens, as this could damage the dura mater and underlying brain tissue. In this case applying a drop of aCSF may help.

   b. Remove any remaining pieces of skull from the craniotomy with the forceps, while avoiding touching the brain surface. Make sure that the edges of the craniotomy are smooth without sharp skull pieces.
      i. Use aCSF to keep the dura mater moist and use absorbent triangles to clean the skull around the craniotomy.
   c. Use a 30G needle mounted on a cotton swab to make a shallow cut in the dura.
      i. Make a cut in the dura on the far right side of the craniotomy.
      ii. Before proceeding make sure that there is no bleeding from ruptured veins in the dura or in the brain. Use repeated application of aCSF and removal of the fluid using absorbent triangles to stop bleedings and keep the brain/dura clean.
      iii. Position one tip of a fine curved forceps (#7 curved) under the dura, avoiding any nearby blood vessels, and carefully remove the dura in one fluent motion.

**CRITICAL: Keep the brain tissue moist with aCSF at all times after removing the dura mater.**

**Note:** The dura mater may be attached to the underlying brain tissue, which could cause damage when the dura is removed too quickly. In such cases, keep the dura well hydrated with aCSF and advance removal of the dura in small steps.

**Optional:** Create a well using UV-curing Optibond primer and dental composite around the craniotomy and fill with aCSF to prevent dehydration of the brain tissue.

**Pause point**: After exposing the brain tissue a break can be taken as long as the brain tissue stays moist. However, because of the reduction in brain tissue temperature and the slow reduction in health of the brain following a craniotomy it is recommended to directly continue with whole-cell recordings.

## Whole-cell recordings with visual stimulation

**Timing: 3-4 h**

In this section whole-cell recordings will be performed targeting neurons in the SC. Visual responses will be measured in current-clamp mode, which allows for characterization of visually-evoked excitatory postsynaptic potentials (EPSPs). While the coordinates indicated in this protocol are aimed at the SC, it is recommended to establish that recordings are indeed made from SC neurons. Inspecting the location of biocytin-filled neurons *post-hoc* will indicate where the targeted neurons are located.

**Note:** There are different visual stimulation techniques available to measure visual processing in SC cells. Most studies use a monitor positioned in front of a mouse to present visual stimuli. For binocular stimulation experiments we use two methods. Firstly, calibrated brief full-field LED flashes can be used to activate retinal ganglion cells (Figure 2A). Two LEDs can be attached to

a moveable arm so that they can be placed on top of the eyes. Black tape can be wrapped around the LEDs and create opaque 'goggles'. This allows for full-field stimulation of each eye individually. Secondly, more complex visual stimuli can be projected on the eyes via monitors placed on either side of the mouse (Figure 2B). Haploscope mirrors are placed in front of the mouse, reflecting the visual images to each eye individually. This method enables a vast range of visual stimulus protocols, including full-screen flashes, moving gratings, moving stimuli. Importantly, stimuli can be presented directly in the receptive field of the cell.

5. Prepare the recording and visual stimulation equipment.
    a. Position a reference electrode (a chloride-coated silver wire) under the neck skin at the back of the incision.
    b. Prepare the visual stimuli (choose one):
        i. Apply continuous voltages to the LEDs so that they emit dim light. Place the LEDs on top of the eyes, making sure that the light illumination is restricted to the eyes only.
        ii. Place the haploscope mirrors in front of the mouse with the snout lightly touching the edge. A laser pointer can be used to check whether the position of the mirrors correctly reflects the visual images on the screens to the eyes.

6. Prepare the recording electrode.
    a. Back-fill an electrode with ICS and attach the electrode to the pipette holder.
    b. Establish high positive pressure, measured by a barometric pressure meter connected to the air tubing system. We establish this by fully depressing the plunger of a 12 mL syringe connected to the system.

7. Position the recording electrode above the SC.
    a. With the patch-clamp amplifier in 'Voltage Clamp', run a continuous 'test-pulse' protocol (*e.g.,* 50 ms ON/OFF, 10 mV step).
    b. Remove the aCSF from the brain tissue using absorbent triangles.
    c. With the micromanipulator set to the coarse speed setting, position the electrode to ~300 $\mu$m above the brain surface.
    d. Adjust the micromanipulator to the fine speed setting and advance the electrode vertically downward until the tip reaches the surface.
    e. Set the micromanipulator coordinates to '0'.

**Note:** We use an Axon MultiClamp 700A patch-clamp amplifier (Molecular Devices) connected to an ITC-18 digitizer (Instrutech). Current and voltage signals are recorded on a Macintosh computer running Axograph at 20 kHz. A MP-285 micromanipulator (Sutter Instrument) was used to control the electrode position.

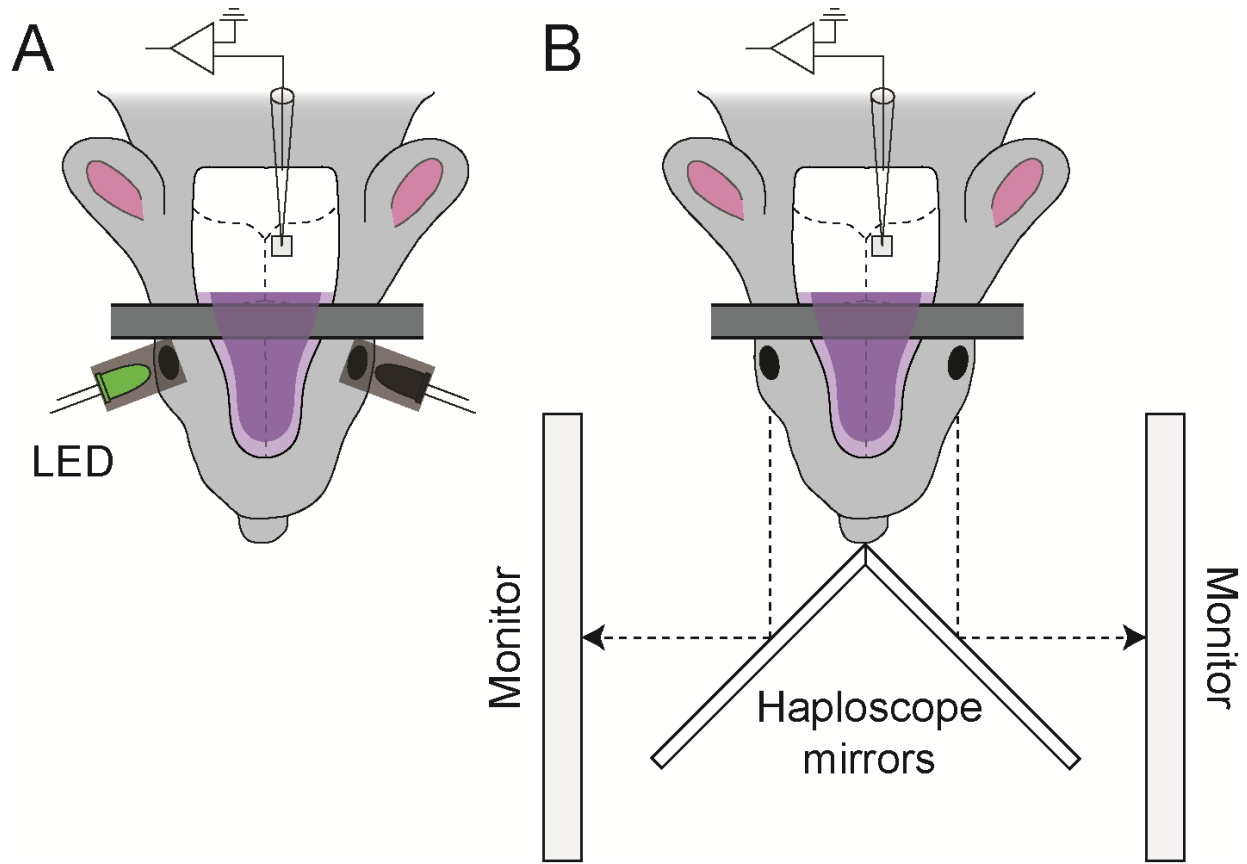

**Figure 2.** Methods for visual stimulation in head-fixed, anesthetized mice. (A) Full-field light flashes are delivered to the eyes using LEDs. LEDs are placed on top of the eyes and enable stimulation of each eye individually. (B) Visual stimulation using a haploscope and monitors on each side enable more complex visual stimuli to the eyes independently.

**Note:** Contact between the electrode and the brain surface can be detected as a sudden increase in resistance as visualized by the data acquisition software. Right before the electrode tip touches the surface, monitor the computer screen to determine when contact is made.

- f. Advance the electrode vertically (~25 μm/s speed) through the tissue until a depth of 200 μm.
- g. Rehydrate the brain tissue by placing a drop of aCSF on the brain surface. Evaluate the quality of the electrode. **[Troubleshooting 3](#)**
    - i. Measure the electrode resistance and replace the electrode if the resistance is lower than 4 MΩ or higher than 7 MΩ.

**Note:** We have observed that electrodes with resistances between 5.5 and 7 MΩ are most effective for *in vivo* recordings from neurons in the SC.

- h. Advance the electrode to a depth of ~1000 μm (~25 μm/s speed) and adjust the pressure to ~20 mbar. We use the mouth and a P20 pipette tip attached to the air tubing system to establish positive pressure.

**Note:** The exact depth of the dorsal surface of the SC is dependent on the anteroposterior and mediolateral position of the electrode, and the degree to which the brain tissue is displaced by the advancing electrode. We have observed that at recording coordinates of 0.8 mm anterior to Lambda and 0.7 mm lateral to midline, the dorsal surface of SC is located at a depth of ~1000 μm.

8. Search for neurons (Figure 3A). **[Troubleshooting 4](#)**
    a. Advance the electrode in steps of 2 μm (2-4 μm/s) while monitoring changes in current. Frequently adjust the pipette offset to zero and keep the pressure at ~20 mbar.
    b. The proximity of a neuron is typically indicated by a sharp and sudden >50% drop in the test pulse current amplitude, reflecting a sudden increase in resistance. The experimenter also often observes a certain 'vertical wobble' in the test pulse current amplitude.
    c. Establish a gigaseal connection.
        i. As soon as above indications are observed, advance another 2-4 μm and quickly release positive pressure and apply mild negative pressure (less than -15 mbar).
        ii. Set the holding voltage of -70 mV early on, as this helps to establish the gigaseal, and gradually release the negative pressure as the resistance approaches 1 GΩ.

**CRITICAL: An ideal gigaseal should form within a few seconds after releasing pressure and starting suction. The time it takes to form the gigaseal is an indication of the strength of suction needed to go whole-cell in the subsequent steps. Fast gigaseal formations generally open at lower suction strengths than those that take >10 seconds. Discard neurons that do not reach a gigaseal or the gigaseal takes a very long to form (>30 sec) as this may indicate debris between the cell membrane and the pipette tip and these recordings will typically have a high access resistance.**

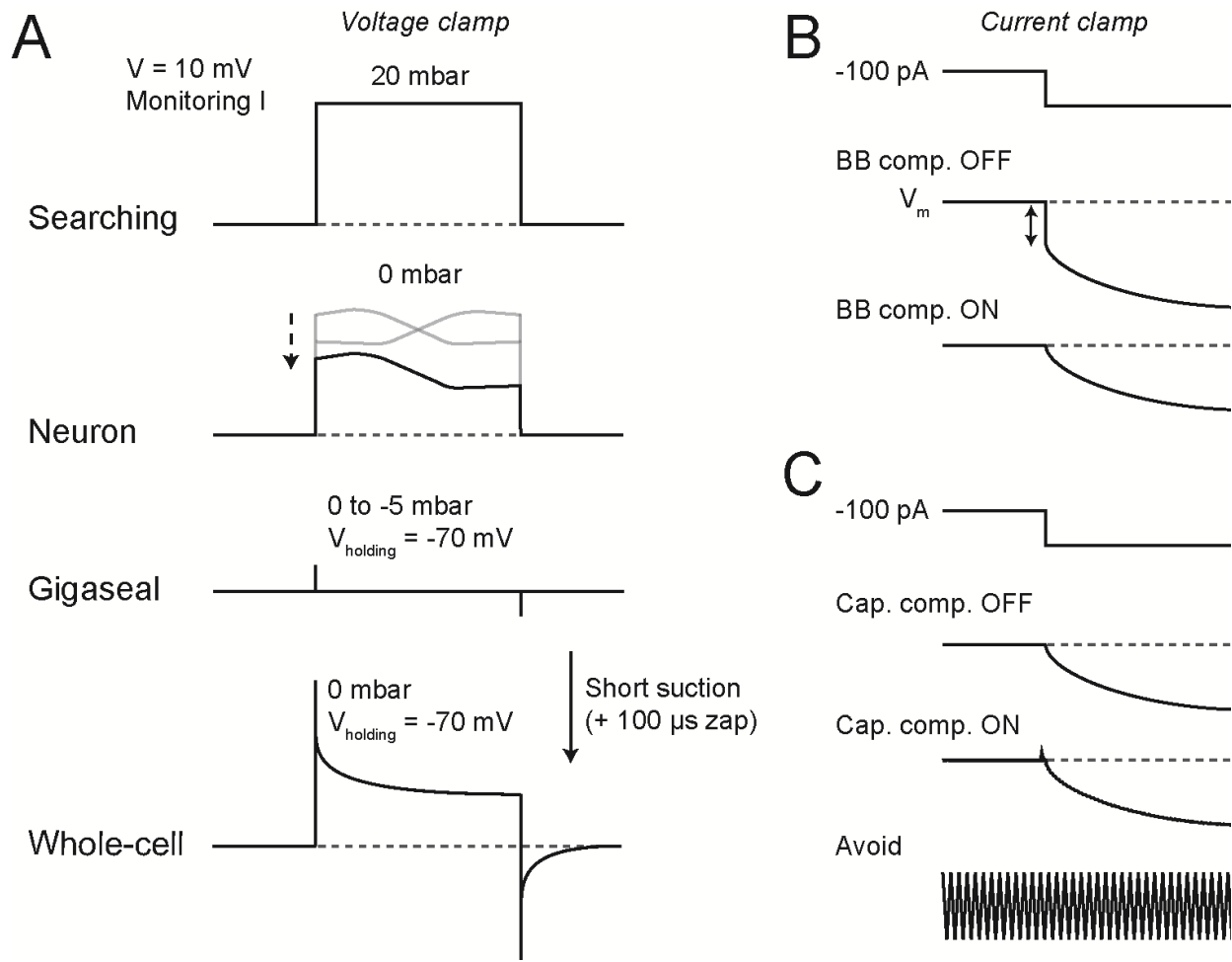

**Figure 3.** Current and voltage responses during different stages of whole-cell recording. (A) Searching for neurons is done in voltage-clamp mode, while monitoring the current during a test pulse. Proximity of a neuron is evident by an increase in resistance at the electrode tip, visible as a reduction of test pulse current. Removal of pressure and applying a holding voltage of -70 mV leads to gigaseal formation. Brief suction pulses (combined with a short zap) break the neuronal membrane at the electrode tip, after which the whole-cell configuration is reached. (B) Negative current pulses in current-clamp mode are used to determine the compensation for the voltage drop due to series resistance (arrows) using bridge balance control. (C) Active current is injected to compensate for the capacitance in the input circuit and electrode using capacitance compensation. Small transients will become visible in the trace at the start of the negative current injection. Great care should be taken to avoid overcompensation of capacitance, which will induce oscillation and can lead to losing the recording.

iii. If no gigaseal is formed replace the patch electrode and repeat step 6-8. Do not attempt to record from other neurons with the same electrode, as once the positive pressure is released the electrode tip will become contaminated.
iv. If a gigaseal forms, compensate for the fast and slow capacitance transients in the cell-attached configuration.

9. Go into the whole-cell configuration. **Troubleshooting 5**
    a. Apply a short pulse of negative suction either by mouth or using a syringe to rupture the cell membrane. A successful 'break-in' can be witnessed by the occurrence of large capacitance transients. (Figure 3A, bottom). Start with gentle suction pulses and increase as required.

**Optional:** If it is not possible to break in using suction alone, suction pulses can be combined with use of the 'Zap' function (1 V DC current for 100 $\mu$s) to rupture the membrane.

**Optional:** Record a voltage step protocol (*e.g.,* 100 ms ON/OFF, 10 mV step) to determine the access resistance and cell's membrane capacitance values.

10. Optimize amplifier settings and characterize the neuron in current clamp. **Troubleshooting 6**
    a. Set the patch-clamp amplifier to 'Current-Clamp' and run a negative current step protocol (*e.g.,* 10 ms ON/OFF, -100 pA) or use the 'Tuning' controls.
    b. Compensate for the voltage drop due to series resistance by adjusting the 'Bridge Balance' controls (Figure 3B).
    c. Compensate for the capacitance of the amplifier and electrode by applying 'Capacitance Neutralisation'. Increase until an overshoot at the onset of the voltage change to a current pulse becomes visible (Figure 3C). Avoid overcompensating which will lead to oscillations and potentially the loss of the recording.
    d. Characterize the voltage response of the cell using a current step protocol comprising of both negative and positive current injections (*e.g.,* 500 ms ON/OFF, -200 pA to +600 pA with 50 pA steps). This will allow cell type classification based electrophysiological parameters, such as input resistance and action potential shape, which differ between cell types (Gale and Murphy, 2014).

11. Record responses to visual stimulation.
    a. Record neuronal responses during alternating visual stimulation of the left, right or both eyes to study visually evoked synaptic inputs.
        i. For LED stimulation we use 20 ms flashes with a 2980 ms inter-trial interval (ITI) with 50 repetitions per stimulus condition.
        ii. For visual stimulation using the haploscope and monitors we use 1.5 s stimulation with a 2.5 s ITI for 50 repetitions per stimulus condition.

**CRITICAL: Allow for sufficient time between stimuli to avoid responses being influenced by preceding stimuli.**

**Note:** Passive diffusion of biocytin during recordings allows for visualization of neuronal morphology after the experiment. Multiple neurons can be recorded in the same hemisphere/animal, but this may complicate correct identification after *post-hoc* labeling of

biocytin. If recording from multiple neurons in the same hemisphere, we recommend leaving sufficient distance in the mediolateral axis between recording sites (>200 μm).

## Transcardial perfusion and removing the brain

**Timing: 30 min**

At the end of the experiment the mouse is transcardially perfused with fixative and the brain is extracted. The fixed tissue is then sliced and processed to allow visualization of the morphology of biocytin-filled cells.

12. Prepare the perfusion setup.
    a. Fill one 50 mL syringe with 30 mL 1X PBS.
    b. Fill one 50 mL syringe with 30 mL 4% PFA solution.
    c. Place the following tools and consumables by the perfusion area:
        i. Large scissors, dissection scissors, spatula, forceps (dumont #3), Graefe forceps, hemostat, Castroviejo scissors, drip tray with cork dissection board, four 25G needles, tissues and a butterfly needle.
        ii. Fill a 15 mL Falcon tube with 4% PFA solution in which to put the brain for post-fixation.
    d. Connect the butterfly needle and the syringe with PBS. Rinse the tubing with PBS and make sure no air bubbles are present.

13. Anesthetize the mouse with an overdose of pentobarbitone solution.
    a. Inject pentobarbitone solution (150 mg/kg) intraperitoneally.
        i. Volume to inject (diluted pentobarbitone solution; 60 mg/mL) = weight of mouse in grams x 0.0025 mL. Inject 0.05 mL of pentobarbitone solution in a 20 g mouse.
    b. Assess the depth of anesthesia using the paw withdrawal reflex after 3 min. Inject additional pentobarbitone solution if the withdrawal reflex persists.

**Note:** In case the mouse stops breathing, immediately perform the paw withdrawal reflex and proceed with the perfusion. The perfusion is preferably carried out while the heart is still beating as this aids the distribution of fixative through the body.

14. Perform transcardial perfusion.
    a. Lie the mouse on its back and secure the limbs in an extended position by pinning to a dissection board with 25-gauge needle tips.
    b. Open the abdominal wall below the rib cage using forceps and dissection scissors.
    c. Carefully cut open the diaphragm in a lateral direction and cut upward through the ribs on both sides. Be careful not to rupture/damage the heart.
    d. Attach the hemostat to the skin flap and place above the mouse to keep the rib cage open.
    e. Expose the heart and insert the butterfly needle into the left ventricle and make a cut in the right atrium with the Castroviejo scissors.
    f. Exert moderate pressure on the plunger to rinse the heart and blood vessels with PBS.

**CRITICAL: A quick color change of the liver from red to gray indicates successful perfusion. If the lungs fill up and fluid comes out of the snout, the needle has penetrated**

**the right ventricle or the cut in the right atrium is too small. In this case reinsert the needle into the left ventricle or make a new cut in the right atrium.**

      g. After the syringe with PBS is empty connect the syringe with 4% PFA and continue the perfusion.
      h. Assess the stiffness of the tail and front paws after the perfusion has been completed. Successful perfusion results in stiff limbs.

**CRITICAL: PFA is toxic. Work in a laboratory safety cabinet and wear appropriate PPE.**

**Note:** Effective perfusion with fresh 4% PFA solution will lead to general muscle contractions.

15. Extract the brain. **[Troubleshooting 7](#)**
    a. Remove the needles that hold the mouse to the dissecting board.
    b. Separate the head and the body using large scissors.
    c. Cut through the skin from the neck until the snout and pull the skin to the side.
    d. Position the tips of the dissection scissors in the eye sockets and cut through the bone between the eyes.
    e. Hold the head so that the eyes are facing the researcher.
    f. Insert one tip of the scissors in the cut between the eye socks. Cut through the skull via the left lateral side of the head in an anterior to posterior direction, while maintaining a dorsal position to avoid damaging the brain.
    g. Use the Graefe forceps to remove the skull by inserting the forceps under the left side of the skull and pulling the entire bone in one motion over to the right side of the head. Remove remaining pieces of skull and loose pieces of dura mater.
    h. Carefully position a spatula under the brain and lift the brain out of the skull base while inverting the skull.
    i. Cut through the optic nerves using the dissection scissors and place the brain in a Falcon tube containing 4% PFA.

**Pause point**: Post-fix the brain between 2 and 24 hours at 4°C. It is recommended that brains are post-fixed for 24 hours if the perfusion was not optimally performed, often visible as filled blood vessels in the brain tissue.

16. Prepare the brain for sectioning.
    a. Place the brain in a 15 mL Falcon tube containing 30% sucrose (w/v) in 1X PBS solution and store at 4°C. Wait until the brain fully sinks usually after 24 hours in sucrose solution.

**Pause point**: Brains in 30% sucrose can be stored long-term in -20°C.

## Preparing coronal sections and immunostaining

**Timing: 30 min**

Fixed brains are processed with a cryostat to create sections. All sections around the recording coordinates will undergo immunostaining and the recorded neuron will be visualized using light microscopy.

17. Slice the brain using a cryostat.
    a. Prepare the cryostat (-20°C).
        i. Secure a razor blade in the holder. We use an angle of 10-15° to cut coronal SC slices. Set the slice thickness to 100 μm.
        ii. Fill a 12-well plate with 1X PBS using a Pasteur pipette.
        iii. Fill a petri dish with a layer of 1X PBS.
    b. Prepare and freeze the brain.
        i. Take the brain out of the 30% sucrose solution using Graefe forceps and place it on a filter paper in a petri dish.
        ii. Use a razor blade to make a cut perpendicular to the horizontal plane to remove the anterior part of the brain. To cut the entire SC, make the cut around -2 mm from Bregma.
        iii. Cut off the cerebellum and brainstem at -5 to -6 mm from Bregma.

**Optional:** Flip the brain forward and make a small cut in the ventral corner on the left or right side to distinguish the left and right hemispheres in the resulting sections.

        iv. Place the brain in a cryomold and immerse in Tissue-Tek Optimal Cutting Temperature (OCT) compound. Place the cryomold on the rapid freezing platform in the cutting chamber (if available) and wait until frozen.
    c. Mount and slice the brain.
        i. Remove the frozen brain from the cryomold. Apply a layer of OCT on the cryostat chuck and place the frozen brain on the chuck. Apply a layer of OCT around the edges of the brain. Place the chuck in the cryostat chamber and wait until frozen.
        ii. Position the chuck firmly onto the moving platform in the cryostat chamber. Position the cutting blade close to the brain and start slicing from the posterior end of the brain forward, until the posterior end of SC is reached around -4.8 mm from Bregma.
        iii. Cut 100 μm coronal brain sections and transport the slices from the cryostat chamber to a 12-well plate using a fine brush. Slice until the anterior end of SC is reached -3 mm to Bregma. Place the slices in order of sectioning.

**Pause point**: Brain slices can be stored in 1X PBS at 4°C for one week.

18. Perform the immunostaining. See materials and equipment for recipes.
    a. We use a Vectastain Elite ABC kit (Vector Laboratories) and perform day 1 of the immunostaining as follows.
        i. Wash sections 5 x 5 min in 1 mL 1X PBS per well in a 12-well plate placed on an orbital shaker.
        ii. Place sections in 1% $H_2O_2$ in 1X PBS for 15 min to block endogenous peroxidase activity.
        iii. Wash sections 3 x 10 min in 1X PBS.
        iv. Place sections in 1% bovine serum albumin (BSA) and 0.3% Triton X-100 in 1X PBS for 1 hour at room temperature.
        v. Incubate sections overnight in ABC solution on an orbital shaker at room temperature.

**Pause point:** Incubate sections overnight in ABC solution on an orbital shaker at room temperature.

b.  Perform day 2 of the immunostaining.
              i.  Wash sections 3 x 5 min in 1 mL 1X PBS per well. Meanwhile, prepare the 3,3'-Diaminobenzidine (DAB) solution as per kit instructions.
              ii. Incubate the sections in the DAB solution for 5-10 min on an orbital shaker at room temperature while monitoring the slices. Keep the slices longer in the solution for darker labeling.
              iii. Wash the slices 3 x 5 min in 1X PBS.
              iv. Mount the slices on microscope slides and cover using curing mounting medium.
        c.  Inspect the slices under a bright-field microscope. **Troubleshooting 8**

**CRITICAL: DAB is toxic. Work in a laboratory safety cabinet and wear appropriate PPE.**

## Expected outcomes

*In vivo* whole-cell recording from neurons in the SC will provide insight into the membrane potential dynamics and spiking activity of SC neurons during visual stimulation of the eyes (Figure 4A). The active and passive properties of recorded neurons can be extracted from their responses to current injection (Figure 4B). The benefit of adding biocytin to the intracellular solution is that it allows for visualizing the neuronal morphology *post hoc* (Figure 4C), aiding cell identification.

Figure 4D shows the membrane potential traces of two example neurons, recorded in current-clamp during visual stimulation with 20 ms LED flashes. There is a large variety of synaptic response patterns to stimulation, consistent with the complexity in the SC network connectivity, input organization and neuronal diversity. We have observed neurons that show excitatory postsynaptic potentials (EPSPs), inhibitory postsynaptic potentials (IPSPs) or combinations of both with a variety of timescales.

A major advantage of whole-cell recordings in vivo is that the entire brain is still intact and therefore the neuron continues to receive active inputs. This is ideal for investigating neuronal encoding in sensory systems that rely on activity in multiple brain areas, such as the visual system. Such systems neuroscience questions cannot be answered with *ex vivo* techniques such as whole-cell recordings from brain slices. In addition, by combining electrophysiological with cell morphology allows for investigating structure-function relations in the SC circuit.

A successful experiment performed by a trained experimenter can be expected to result in good quality data from 1-3 neurons per preparation, but we find that the success rate of this technique is highly variable. A particularly important factor that determines success is the quality of the brain tissue after the craniotomy. The experimenter should take great care to prevent damage to the brain tissue when creating the craniotomy and removing the dura.

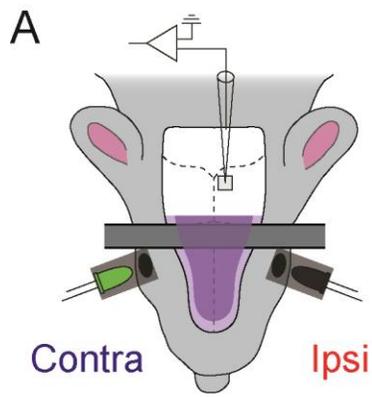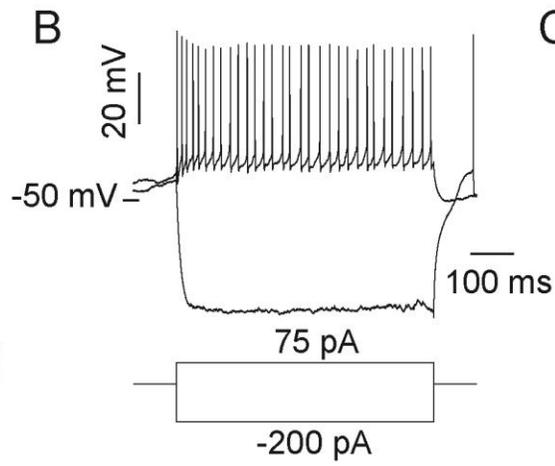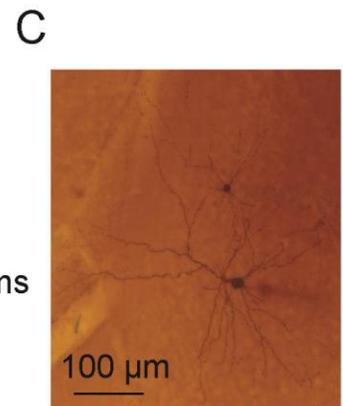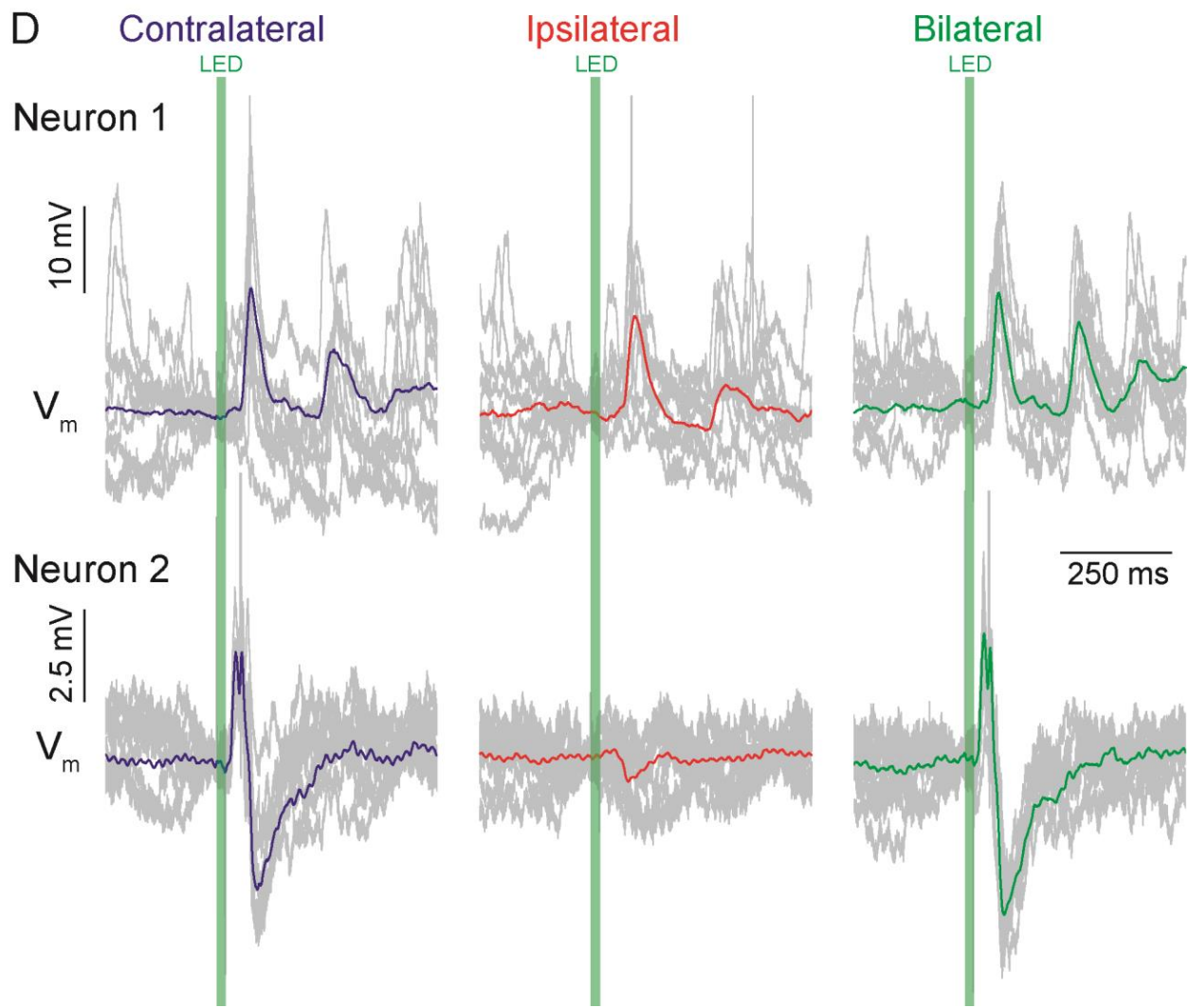

**Figure 4.** Example of *in vivo* whole-cell recordings during LED stimulation of the eyes. (A) Schematic overview of the experimental setup for making recordings from superior colliculus neurons. (B) Electrophysiological properties can be characterized by a neuron's response to negative and positive current steps. (C) Example of two biocytin-filled neurons in SC visualized using DAB staining. (D) Membrane potential dynamics of two example neurons during stimulation of the contralateral or ipsilateral eye, or during bilateral stimulation. The timing of 20 ms LED flashes is indicated in green. Ten raw traces are plotted (grey) with the average superimposed. Spikes are trimmed and no holding current was injected.

## Limitations

Although this protocol provides a method for obtaining quality recordings with high temporal resolution, experimenters should be aware of the limitations. Recordings from deeper brain structures, such as the anteromedial part of the SC where binocular neurons are located, involves penetration of overlaying cortical brain tissue. This will cause damage that could affect neuronal activity and disrupt physiological functions. In particular, to reach the anteromedial SC electrodes have to cross through the retrosplenial cortex, which has connections with the SC and evidence has emerged showing that this pathways plays a role in refuge localization (Vale et al., 2020). The posterior regions of SC have less or no overlaying cortical tissue and recording from these regions is preferred if this suits the study design. Moreover, experimenters that are learning to perform this technique are advised to target the posterior SC first before targeting the subcortical part of SC.

One consequence of needing to penetrate through brain tissue is that the electrode tip may become contaminated, leading to difficulties obtaining a gigaseal. In addition, blood vessels may be pierced which could result in internal bleedings. Experimenters should always inspect the electrode tip for blood traces when retracting. If there are indications of bleeding, the next electrode should be inserted >100 μm away from the previous site. We recommend that the number of electrode insertions is kept as low as possible to limit tissue damage. In our experience, if no neurons were recorded after ~10 electrode insertions, it is best to patch in a different brain region or stop the experiment. We have also found that it becomes more difficult to obtain recordings in older mice and we therefore recommend starting using animals with an age of 4-6 weeks. Experimenters should be aware that all of these complications together increase the difficulty of obtaining SC recordings. We have found that both the quality and the yield of whole-cell recordings in SC is lower, for example, than in the primary visual cortex (V1).

Another limitation is that in order to study morphology one should avoid recording from multiple neurons in the same area to avoid confusion with cell identification. One way to do this is to record from only 1 neuron per hemisphere. This limits the data per animal and results in more animals having to be used for a complete dataset. Furthermore, since recordings are made blindly it is often not clear from which SC layer and cell-type the recording has been made. This can be determined *post-hoc* based on the neuron's voltage responses during current step protocols as well as their morphology (Gale and Murphy, 2014, 2016, 2018). Another important limitation is that general anesthesia has a strong effect on SC physiology and its responses to visual stimulation (De Franceschi and Solomon, 2018; Kasai and Isa, 2022; Populin, 2005). This complicates extrapolating responses in anesthetized animals to the awake, physiological state. While urethane-CPX anesthesia is an often-used anesthetic, experimenters may want to consider using other anesthetics such as ketamine-xylazine or isoflurane, depending on the requirements of the investigation.

When recording from an intact brain, it is often unclear what underlies visually-evoked as well as spontaneous transients. As can be observed in Figure 4D, visual stimulation can result in multiple 'waves' of synaptic activity. Determining the circuits responsible for these responses can be challenging. It is also questionable whether synaptic activity with longer latencies still code visual information, or whether they are the result of synchronized, generalized activity in multiple (visual and non-visual) brain areas. Furthermore, while drugs may be easily added to the bath solution during *in vitro* slice recordings, such an approach is challenging *in vivo*, particularly when recording from deeper brain structures. Three potential solutions are 1) adding the drugs to the ICS, although due to the use of positive pressure to keep the electrode tip uncontaminated the drugs will also be ejected from the electrode and therefore present in the extracellular space. This procedure also does not allow for recording responses without drugs; 2) topical application of drugs on the brain surface. This procedure is suitable for superficial recordings, but not for recordings deeper in the brain; and 3) adopting a pipette-pair approach enabling simultaneous recordings and local pharmacology (*e.g.,* Kheradpezhouh et al., 2021). The experimenter will have to weigh the pros and cons of each approach.

Lastly, there are complications in making direct comparisons between *in vitro* recordings from brain slices and *in vivo* whole-cell recordings from anesthetized rodents. *In vivo* recordings usually have a higher access resistance, neurons can be more depolarized and there may be more active synaptic inputs as a result of an intact neural circuit.

# Troubleshooting

## Problem 1:

Mouse keeps showing the paw withdrawal reflex after injecting the anesthetic (step 1).

Mouse has a low or irregular breathing rhythm (any time during the experiment).

## Potential solution:

The concentration of the anesthetic may be too low. Check whether the preparation of the urethane-CPX mixture has been done correctly and prepare a new mixture if necessary. Alternatively, the intraperitoneal injection may not have been done correctly, leading to injection of the mixture into organs or muscle. In this case the anesthesia takes longer to take effect.

Injecting a higher dose than recommended can result in irregular breathing or a lower breathing rhythm. Check whether the preparation of the urethane-CPX mixture has been done correctly and prepare a new mixture if necessary.

## Problem 2:

Head bar detaches from the skull (step 2).

## Potential solution:

The combination of primer and dental composite usually forms a very strong attachment of the head bar to the skull. Check if all the steps have been followed correctly. One potential issue is that the skull may not have dried sufficiently before applying the primer/composite. Bleeding between the skull and the primer layer may also have occurred. For solid attachment it is important to dry the skull completely and to stop any bleeding.

### Problem 3:

Pipette resistance is high (>7 MΩ) and does not decrease even under high air pressure (step 7g).

### Potential solution:

If the pipette resistance is within acceptable range when touching the brain tissue (between 4 and 7 MΩ), but increases when moving the tip to the SC, this could indicate that the electrode has hit a blood vessel, which can cause internal bleeding. Blood may contaminate or block the electrode tip, leading to a higher pipette resistance. An indication that a bleeding is increasing the pipette resistance is that applying high pressure temporarily reduces the resistance after which it increases again. If this occurs a new electrode should be used to target another region of SC.

### Problem 4:

Difficulties finding neurons or establishing gigaseals. Changes in resistance indicating the presence of neurons (Figure 3A) is not observed (or only rarely) when progressing in a step-wise fashion through the tissue (step 8).

### Potential solution:

There could be several reasons for this. First, it is important to check whether the pipette tip is targeting the correct brain region, since variations in head bar placement result in different angles of approach. Second, it could be that the targeted area contains a low density of healthy neurons through cell death. A new electrode should be used to target an adjacent region of SC. Third, internal bleeding can lead to contamination of the electrode tip, making it more difficult to find neurons and establish gigaseals. When retracting the electrode check the tip for indications of bleeding. A new electrode should be used to target a different region of SC.

### Problem 5:

Neuron closes soon after break-in (<5 min). Whole-cell configuration cannot be maintained for long time duration (step 9 and 10).

### Potential solution:

This usually indicates that the connection between the electrode tip and the neuron has not been sufficiently opened. Sufficiently strong air suction should be applied to open the connection when going whole-cell. During recordings mild positive or negative pressure can help maintain the open connection. Alternatively, there could be movement of the electrode or brain tissue that affects the connection. Check whether the head bar is secure and that the animal is stable. Also check that the electrode is securely held by the pipette holder and that this is securely attached to the amplifier headstage.

## Problem 6:

There are high (50-60 Hz), medium (5-15 Hz) or low (1-4 Hz) frequency oscillations in the voltage trace after break-in (step 10).

Resting membrane potential is depolarized after break-in or rises during the recording (step 10).

The neuron continuously spikes after break-in (step 10).

## Potential solution:

High frequency oscillations (50-60 Hz, depending on geographic location) are generally caused by electrical noise in the recording faraday cage. Check whether all components are well grounded. Disconnect electrical components and connect them one-by-one to identify the source of the electrical noise. Also check whether the reference electrode is placed firmly under the skin and that the skin is moist. Re-chloride the reference and electrode silver-wire if necessary. Medium (5-15 Hz) or low (1-4 Hz) frequency oscillations generally indicate movement of the electrode or brain tissue caused by the mouse's heartbeat or breathing, respectively. Make sure all components are well secured and are not moving. Oscillations in the trace can also indicate an imperfect connection between the electrode tip and the neuron or that debris is blocking that connection. A short change in air pressure or slightly retracting the electrode may fix the issue.

A depolarized resting membrane potential after break-in generally indicates an unhealthy neuron and/or bad quality recording. Discard the recording if the resting membrane potential is higher than ~45 mV, although some cell types may be more depolarized at rest. A slowly rising membrane potential could indicate bad ICS. Check whether the ICS has been prepared correctly and prepare new ICS if necessary.

A spiking neuron after break-in could indicate a bad quality recording or unhealthy neuron, especially when coinciding with a depolarized resting membrane potential. That said, neurons are generally more active *in vivo* than when recorded *in vitro*. Furthermore, some neurons exhibit high spontaneous spike rates even in slices, such as cerebellar Purkinje cells (Llinás and Sugimori, 1980).

## Problem 7:

The brain tissue is red/pink and blood in blood vessels is visible after extracting the brain (step 15).

## Potential solution:

This indicates that the perfusion has not gone well. The perfusion has likely not led to sufficient PFA to reach all parts of the body. For future perfusions make sure that the butterfly needle is correctly positioned in the left heart chamber. If fluid exists the body through the snout or mouth it is best to reinsert the needle.

## Problem 8:

Difficulties finding recorded neurons in the DAB-treated brain slices (step 18c).

### Potential solution:

There could be several reasons for this issue. First, the DAB staining may have not been successfully performed. Check that all the steps have been performed correctly. Second, biocytin filling of the neuron may have been unsuccessful. Longer recordings with lower access resistance generally result in better biocyin labeling. Ideally, recordings should last at least 20 minutes. The mouse should also be kept alive for at least 30 minutes after the recording to allow sufficient diffusion of biocytin through all parts of the neuron.

## Resource availability

**Lead contact**

Further information and requests for resources and reagents should be directed to and will be fulfilled by the lead contact, Robin Broersen (r.broersen@erasmusmc.nl)

**Materials availability**

This study did not generate new unique reagents.

**Data and code availability**

The datasets and code supporting the current study have not been deposited in a public repository because they will soon be published as part of another study. Data and code is however available from the corresponding author on request.

## Acknowledgments

This work was supported by the Australian Research Council Centre of Excellence for Integrative Brain Function (ARC Centre Grant CE140100007). We thank Dr. H. Huang and Dr. W.M. Connelly for their contributions to the study.

## Author contributions

R.B. performed the experiments, analyzed the data and wrote the manuscript. G.J.S. wrote and edited the manuscript and acquired the funding.

## Declaration of interests

The authors declare no competing interests.

# References

De Franceschi, G., and Solomon, S.G. (2018). Visual response properties of neurons in the superficial layers of the superior colliculus of awake mouse. J Physiol *596*, 6307–6332. https://doi.org/10.1113/JP276964.

Gale, S.D., and Murphy, G.J. (2014). Distinct Representation and Distribution of Visual Information by Specific Cell Types in Mouse Superficial Superior Colliculus. J. Neurophysiol. *34*, 13458–13471. https://doi.org/10.1523/JNEUROSCI.2768-14.2014.

Gale, S.D., and Murphy, G.J. (2016). Active Dendritic Properties and Local Inhibitory Input Enable Selectivity for Object Motion in Mouse Superior Colliculus Neurons. J. Neurophysiol. *36*, 9111–9123. https://doi.org/10.1523/JNEUROSCI.0645-16.2016.

Gale, S.D., and Murphy, G.J. (2018). Distinct cell types in the superficial superior colliculus project to the dorsal lateral geniculate and lateral posterior thalamic nuclei. J. Neurophysiol. *120*, 1286–1292. https://doi.org/10.1152/jn.00248.2018.

Haider, B., Schulz, D.P.A., Häusser, M., and Carandini, M. (2016). Millisecond Coupling of Local Field Potentials to Synaptic Currents in the Awake Visual Cortex. Neuron *90*, 35–42. https://doi.org/10.1016/j.neuron.2016.02.034.

Kasai, M., and Isa, T. (2022). Effects of Light Isoflurane Anesthesia on Organization of Direction and Orientation Selectivity in the Superficial Layer of the Mouse Superior Colliculus. J. Neurosci. *42*, 619–630. https://doi.org/10.1523/JNEUROSCI.1196-21.2021.

Kheradpezhouh, E., Mishra, W., and Arabzadeh, E. (2021). A protocol for simultaneous in vivo juxtacellular electrophysiology and local pharmacological manipulation in mouse cortex. STAR Protocols *2*, 100317. https://doi.org/10.1016/j.xpro.2021.100317.

Llinás, R., and Sugimori, M. (1980). Electrophysiological properties of in vitro Purkinje cell somata in mammalian cerebellar slices. J Physiol *305*, 171–195. https://doi.org/10.1113/jphysiol.1980.sp013357.

Longordo, F., To, M.-S., Ikeda, K., and Stuart, G.J. (2013). Sublinear integration underlies binocular processing in primary visual cortex. Nature Neuroscience *16*, 714–723. https://doi.org/10.1038/nn.3394.

Populin, L.C. (2005). Anesthetics Change the Excitation/Inhibition Balance That Governs Sensory Processing in the Cat Superior Colliculus. J. Neurosci. *25*, 5903–5914. https://doi.org/10.1523/JNEUROSCI.1147-05.2005.

Vale, R., Campagner, D., Iordanidou, P., Arocas, O.P., Tan, Y.L., Stempel, A.V., Keshavarzi, S., Petersen, R.S., Margrie, T.W., and Branco, T. (2020). A cortico-collicular circuit for accurate orientation to shelter during escape. BioRxiv 2020.05.26.117598. https://doi.org/10.1101/2020.05.26.117598.